\documentclass[runningheads]{llncs}
\usepackage{xcolor}
\usepackage{verbatim}
\usepackage{hyperref}
\hypersetup{
    colorlinks=true,
    linkcolor=blue,
    filecolor=magenta,      
    urlcolor=blue,
    pdfpagemode=,
    }
\usepackage{makecell}
\usepackage{graphicx}
\usepackage{colortbl}
\usepackage{amssymb}
\usepackage{amsmath}
\usepackage{pifont}

\usepackage{lipsum}
\usepackage[noend]{algpseudocode}
\usepackage[ruled,vlined]{algorithm2e}
\usepackage{rotating}
\usepackage[export]{adjustbox}

\usepackage{subfigure}
\usepackage{dsfont}

\usepackage[
  separate-uncertainty = true,
  multi-part-units = repeat
]{siunitx}

\begin{document}
%



\title{Detecting the Sensing Area of A Laparoscopic Probe in Minimally Invasive Cancer Surgery}

\author{Baoru Huang\inst{1,2} \and Yicheng Hu\inst{1,2} \and
Anh Nguyen\inst{3}  \and
Stamatia Giannarou \inst{1,2}  \and
Daniel S. Elson \inst{1,2}}

\institute{The Hamlyn Centre for Robotic Surgery, Imperial College London, London, UK \\
\email{Baoru.Huang18@imperial.ac.uk} \and
Department of Surgery $\&$ Cancer, Imperial College London, London, UK \and
Department of Computer Science, University of Liverpool, Liverpool, UK 
\\}

\maketitle              
\begin{abstract}
In surgical oncology, it is challenging for surgeons to identify lymph nodes and completely resect cancer even with pre-operative imaging systems like PET and CT, because of the lack of reliable intraoperative visualization tools. Endoscopic radio-guided cancer detection and resection has recently been evaluated whereby a novel tethered laparoscopic gamma detector is used to localize a preoperatively injected radiotracer. This can both enhance the endoscopic imaging and complement preoperative nuclear imaging data. However, gamma activity visualization is challenging to present to the operator because the probe is non-imaging and it does not visibly indicate the activity origination on the tissue surface. Initial failed attempts used segmentation or geometric methods, but led to the discovery that it could be resolved by leveraging high-dimensional image features and probe position information. To demonstrate the effectiveness of this solution, we designed and implemented a simple regression network that successfully addressed the problem. To further validate the proposed solution, we acquired and publicly released two datasets captured using a custom-designed, portable stereo laparoscope system. Through intensive experimentation, we demonstrated that our method can successfully and effectively detect the sensing area, establishing a new performance benchmark. 
Code and data are available at \href{https://github.com/br0202/Sensing_area_detection.git}{https://github.com/br0202/Sensing\_area\_detection.git}.

\end{abstract}
\setcounter{footnote}{0} 
\section{Introduction}
Cancer remains a significant public health challenge worldwide, with a new diagnosis occurring every two minutes in the UK (Cancer Research UK\footnote{\href{https://www.cancerresearchuk.org/health-professional/cancer-statistics-for-the-uk}{https://www.cancerresearchuk.org/health-professional/cancer-statistics-for-the-uk}}). Surgery is one of the main curative treatment options for cancer. However, despite substantial advances in pre-operative imaging such as CT, MRI, or PET/SPECT to aid diagnosis, surgeons still rely on the sense of touch and naked eye to detect cancerous tissues and disease metastases intra-operatively due to the lack of reliable intraoperative visualization tools. In practice, imprecise intraoperative cancer tissue detection and visualization results in missed cancer or the unnecessary removal of healthy tissues, which leads to increased costs and potential harm to the patient. There is a pressing need for more reliable and accurate intraoperative visualization tools for minimally invasive surgery (MIS) to improve surgical outcomes and enhance patient care.

\begin{figure}[t]
\centering
\subfigure[]
{\includegraphics[width=0.49\linewidth, height=0.31\linewidth]{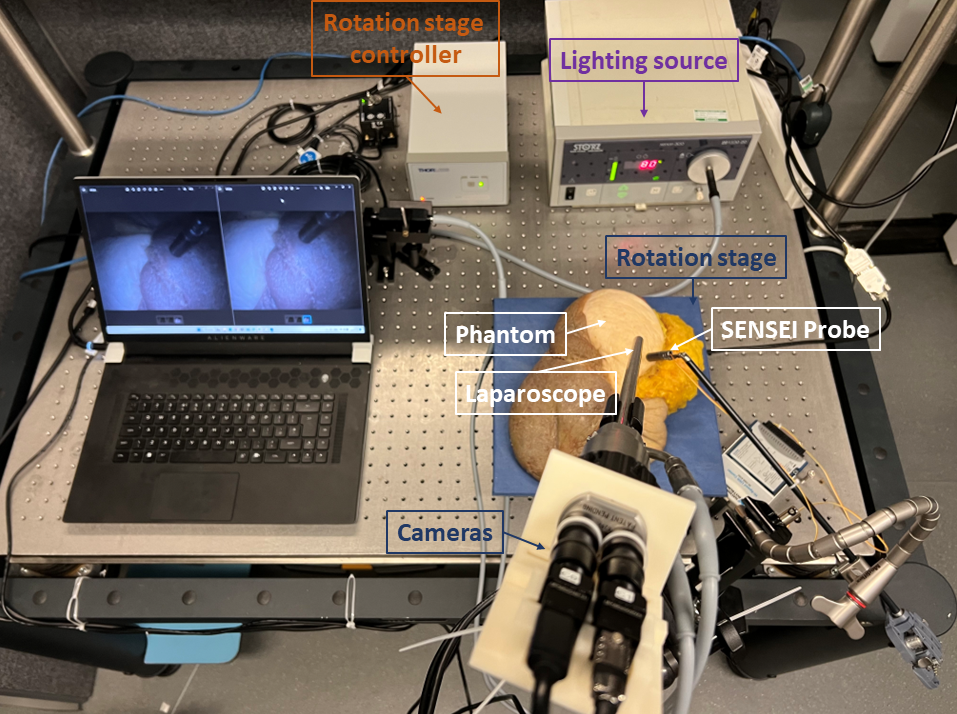}}
\subfigure[]
{\includegraphics[width=0.49\linewidth, height=0.31\linewidth]{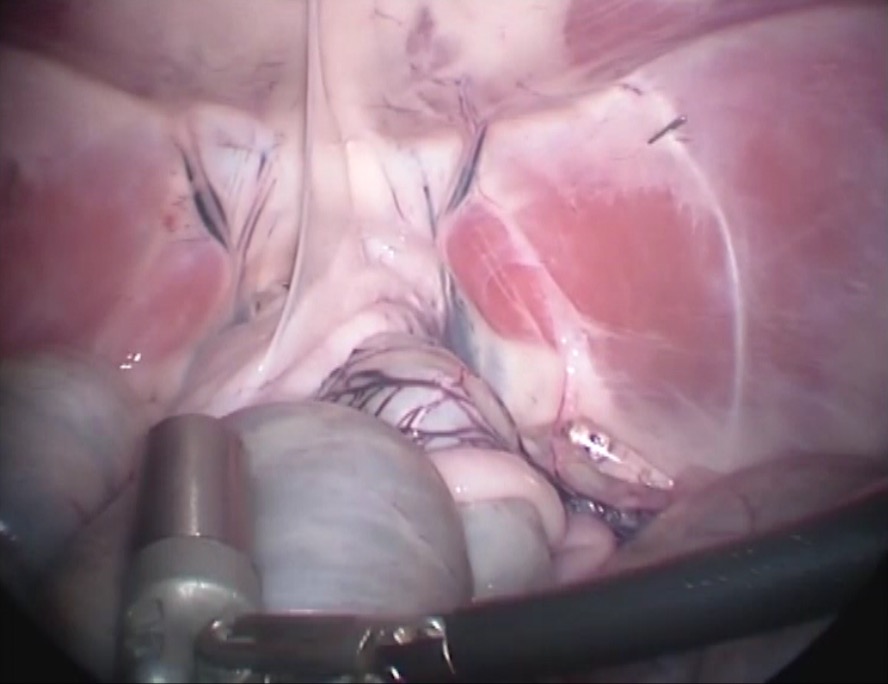}}
\vspace{-2ex}
\caption{(a) Hardware set-up for experiments, including a customized portable stereo laparoscope system and the `SENSEI' probe, a rotation stage, a laparoscopic lighting source, and a phantom; (b) An example of the use of the `SENSEI' probe in MIS.}
\label{fig:intro}
\vspace{-3ex}
\end{figure}


A recent miniaturized cancer detection probe (i.e., `SENSEI\textsuperscript{\textregistered}' developed by Lightpoint Medical Ltd.) leverages the cancer-targeting ability of nuclear agents typically used in nuclear imaging to more accurately identify cancer intra-operatively from the emitted gamma signal (see Fig. 1b)\cite{huang2020tracking}. However, the use of this probe presents a visualization challenge as the probe is non-imaging and is air-gapped from the tissue, making it challenging for the surgeon to locate the probe-sensing area on the tissue surface. 

It is crucial to accurately determine the sensing area, with positive signal potentially indicating cancer or affected lymph nodes. Geometrically, the sensing area is defined as the intersection point between the gamma probe axis and the tissue surface in 3D space, but projected onto the 2D laparoscopic image. However, it is not trivial to determine this using traditional methods due to poor textural definition of tissues and lack of per-pixel ground truth depth data. Similarly, it is also challenging to acquire the probe pose during the surgery. 

\textbf{Problem redefinition.} In this study, in order to provide  sensing area visualization ground truth, we modified a non-functional `SENSEI' probe by adding a miniaturized laser module to clearly optically indicate the sensing area on the laparoscopic images - i.e. the `probe axis-surface intersection'. Our system consists of four main components: a customized stereo laparoscope system for capturing stereo images, a rotation stage for automatic phantom movement, a shutter for illumination control, and a DAQ-controlled switchable laser module (see Fig.~\ref{fig:intro}a). 
With this setup, we aim to transform the sensing area localization problem from a geometrical issue to a high-level content inference problem in 2D. It is noteworthy that this remains a challenging task, as ultimately we need to infer the probe axis-surface intersection without the aid of the laser module to realistically simulate the use of the `SENSEI' probe. 


\section{Related Work}

Laparoscopic images play an important role in computer-assisted surgery and have been used in several problems such as object detection~\cite{jo2019robust}, image segmentation~\cite{yoon2022surgical}, depth estimation ~\cite{tukra2022stereo} or 3D reconstruction~\cite{liu2022sage}. Recently, supervised or unsupervised depth estimation methods have been introduced \cite{liu2019dense}. Ye
\textit{et al.}~\cite{ye2017self} proposed a deep learning framework for surgical scene depth estimation
in self-supervised mode and achieved scalable data acquisition by incorporating a differentiable spatial transformer and an autoencoder into their framework. A 3D displacement module was explored in~\cite{xu2022self} and 3D geometric consistency was utilized in~\cite{huang2022self} for self-supervised monocular depth estimation. Tao~\textit{et al.} \cite{tao2023svt} presented a spatiotemporal vision transformer-based method and a self-supervised generative adversarial network was introduced in~\cite{huang2021self} for depth estimation of stereo laparoscopic images. Recently, fully supervised methods were summarized in~\cite{allan2021stereo} for depth estimation. 
However, acquiring per-pixel ground truth depth data is challenging, especially for laparoscopic images, which makes it difficult for large-scale supervised training~\cite{huang2022self}. 

Laparoscopic segmentation is another important task in computer-assisted surgery as it allows for accurate and efficient identification of instrument position, anatomical structures, and pathological tissue. For instance, a unified framework for depth estimation and surgical tool segmentation in laparoscopic images was proposed in~\cite{huang2022simultaneous}, with simultaneous depth estimation and segmentation map generation. In~\cite{liu2020self}, self-supervised depth estimation was utilized to regularize the semantic segmentation in knee arthroscopy. Marullo~\textit{et al.}~\cite{marullo2023multi} introduced a multi-task convolutional neural network for event detection and semantic segmentation in laparoscopic surgery. The dual swin transformer U-Net was proposed in~\cite{lin2022ds} to enhance the medical image segmentation performance, which leveraged the hierarchical swin transformer into both the encoder and the decoder of the standard U-shaped architecture, benefiting from the self-attention computation in swin transformer as well as the dual-scale encoding design. 

Although the intermediate depth information was not our final aim and can be bypassed, the 3D surface information was necessary in the intersection point inference. ResNet~\cite{he2016deep} has been commonly used as the encoder to extract the image features and geometric information of the scene. In particular, in~\cite{xu2022self}, concatenated stereo image pairs were used as inputs to achieve better results, and such stereo image types are also typical in robot-assisted minimally invasive surgery with stereo laparoscopes. Hence, stereo image data was also adopted in this paper.

If the problem of inferring the intersection point is treated as a geometric problem, both data collection and intra-operative registration would be difficult, which inspired us to approach this problem differently. In practice, we utilize the laser module to collect the ground truth of the intersection points when the laser is on. We note that the standard illumination image from the laparoscopic probe is also captured with the same setup when the laser module is on. Therefore, we can establish a dataset with an image pair (RGB image and laser image) that shares the same intersection point ground truth from the laser image (see Fig.~\ref{fig:dataset}a and Fig.~\ref{fig:dataset}b).
The assumptions made are that the probe's 3D pose when projected into the two 2D images is the observed 2D pose, and that the intersection point is located on its axis. Hence, we input these axes to the network as another branch and randomly sampled points along them to represent the probe.



\section{Dataset}
\begin{figure}[h]
\centering
\includegraphics[width=0.99\linewidth]{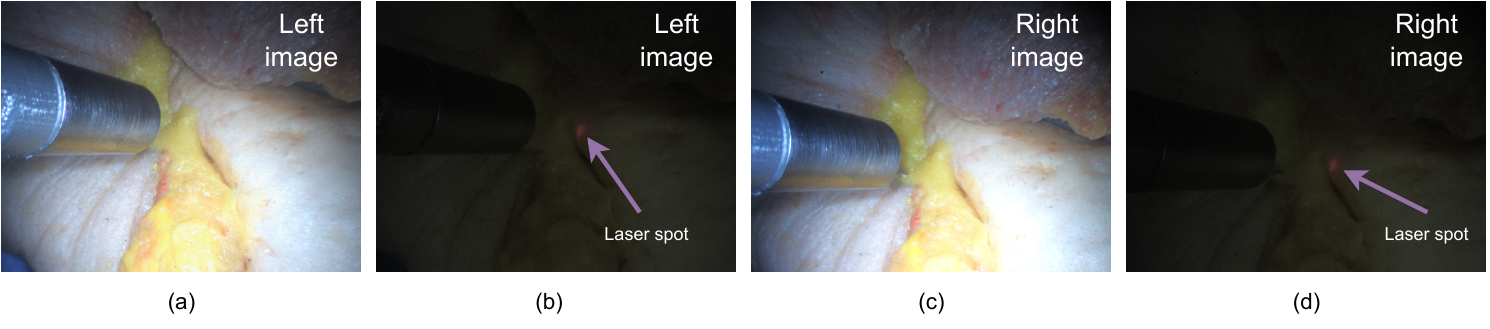}
\vspace{-2ex}
\caption{Example data. (a) Standard illumination left RGB image; (b) left image with laser on and laparoscopic light off; same for (c) and (d) but for right images.}
\label{fig:dataset}
\vspace{-3ex}
\end{figure}

To validate our proposed solution for the newly formulated problem, we acquired and publicly released two new datasets. In this section, we introduce the hardware and software design that was used to achieve our final goal, while Fig.~\ref{fig:dataset} shows a sample from our dataset.

\textbf{Data Collection.}
Two miniaturized, high-resolution cameras were coupled onto a stereo laparoscope using a custom-designed connector. The accompanying API allowed for automatic image acquisition, exposure time adjustment, and white balancing. An electrically controllable shutter was incorporated into the standard laparoscopic illumination path. To indicate the probe axis-surface intersection, we incorporated a DAQ controlled cylindrical miniature laser module into a `SENSEI' probe shell so that the adapted tool was visually identical to the real probe. The laser module emitted a red laser beam (wavelength 650 \textit{nm}) that was visible as a red spot on the tissue surface.

We acquired the dataset on a silicone tissue phantom which was \(30 \times 21 \times 8 \) \textit{cm} and was rendered with tissue color manually by hand to be visually realistic. The  phantom was placed on a rotation stage that stepped $10$ times per revolution to provide views separated by a 36-degree angle. At each position, stereo RGB images were captured \textit{i)} under normal laparoscopic illumination with the laser off; \textit{ii)} with the laparoscopic light blocked and the laser on; and \textit{iii)} with the laparoscopic light blocked and the laser off. Subtraction of the images with laser on and off readily allowed segmentation  of the laser area and calculation of its central point, i.e. the ground truth probe axis-surface intersection.

All data acquisition and devices were controlled by Python and LABVIEW programs, and complete data sets of the above images were collected on visually realistic phantoms for multiple probe and laparoscope positions. This provided 10 tissue surface profiles for a specific camera-probe pose, repeated for 120 different camera-probe poses, mimicking how the probe may be used in practice. Therefore, our first newly acquired  dataset, named \textbf{Jerry}, contains 1200 sets of images. Since it is important to report errors in 3D and in millimeters, we recorded another dataset similar to \textbf{Jerry} but also including ground truth depth map for all frames by using structured-lighting system~\cite{huang2022self} --- namely the \textbf{Coffbee} dataset.

These datasets have multiple uses such as:
\begin{itemize}
    \item Intersection point detection: detecting intersection points is an important problem that can bring accurate surgical cancer visualization. We believe this is an under-investigated problem in surgical vision.
    \item Depth estimation: corresponding ground truth will be released.
    \item Tool segmentation: corresponding ground truth will be released.
\end{itemize}

\section{Probe Axis-Surface Intersection Detection}

\subsection{Overview}


The problem of detecting the intersection point is trivial when the laser is on and can be solved by training a deep segmentation network. However, segmentation requires images with a laser spot as input, while the real gamma probe produces no visible mark and therefore this approach produces inferior results.

An alternative approach to detect the intersection point is to reconstruct the 3D tissue surface and estimate the pose of the probe in real time. A tracking and pose estimation method for the gamma probe \cite{huang2020tracking} involved attaching a dual-pattern marker to the probe to improve detection accuracy. This enabled the derivation of a 6D pose, comprising a rotation matrix and translation matrix with respect to the laparoscope camera coordinate. To obtain the intersection point, the authors used the Structure From Motion (SFM) method to compute the 3D tissue surface, combining it with the estimated pose of the probe, all within the laparoscope coordinate system. However, marker-based tracking and pose estimation methods have sterilization implications for the instrument, and the SFM method requires the surgeon to constantly move the laparoscope, reducing the practicality of these methods for surgery.

In this work, we propose a simple, yet effective regression approach to address this problem. Our approach relies solely on the 2D information and works well without the need for the laser module after training. Furthermore, this simple methodology facilitated an average inference time of $50$ frames per second, enabling real-time sensing area map generation for intraoperative surgery. 

\subsection{Intersection Detection as Segmentation} 
We utilized different deep segmentation networks as a first attempt to address our problem~\cite{ronneberger2015u,koch2015siamese}. Please refer to the Supplementary Material for the implementation details of the networks. We observed that when we do not use images with the laser, the network was not able to make any good predictions. This is understandable as the red laser spot provides the key information for the segmentation. Therefore the network does not have any visual information to make predictions from images of the gamma probe. We note that to enable real-world applications, we need to estimate the intersection point using the images when the laser module is turned off.

\subsection{Intersection Detection as Regression}

\begin{figure}[t]
\centering
\includegraphics[width=0.99\linewidth]{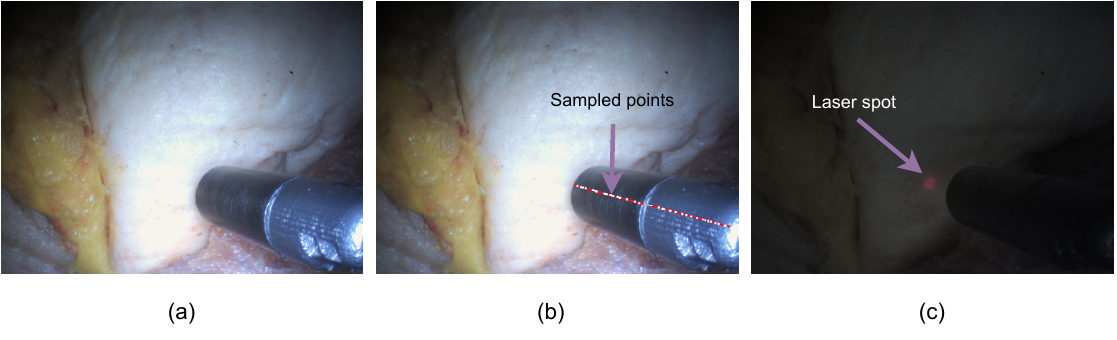}
\vspace{-2ex}
\caption{Sensing area detection. (a) The input RGB image, (b) The estimated line using PCA for obtaining principal points, (c) The image with laser on that we used to detect the intersection ground truth.}
\label{fig_problem_define}
\end{figure}

\begin{figure}[t]
\centering
\includegraphics[width=0.99\linewidth, height=0.35\linewidth]{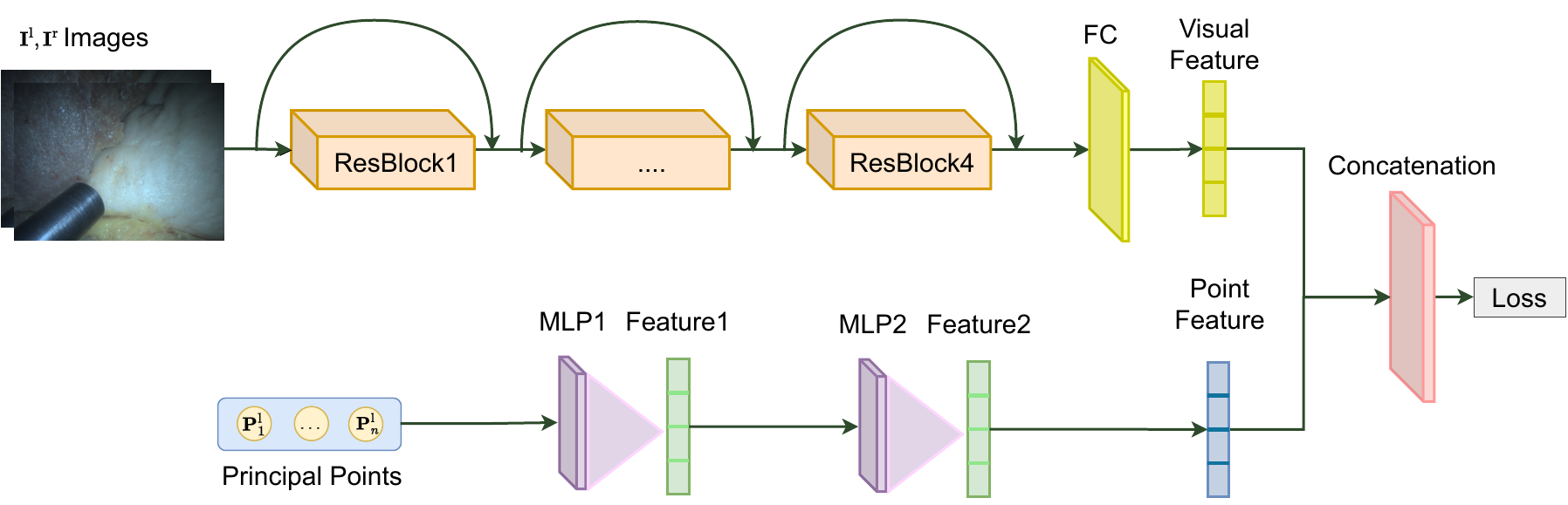}
\vspace{-2ex}
\caption{An overview of our approach using ResNet and MLP. 
}
\label{fig_network}
\end{figure}

\subsubsection{Problem Formulation.} Formally, given a pair of stereo images $\mathbf{I}^{\rm l},\mathbf{I}^{\rm r}$,  $n$ points $\{\mathbf{P}_1^{\rm l}, \mathbf{P}_2^{\rm l}, ...,\mathbf{P}_n^{\rm l}\}$ were sampled along the principal axis of the probe, $\mathbf{P}_i^{\rm l} \in \mathds{R}^{2}$ from the left image. The same process was repeated for the right image. The goal was to predict the intersection point $\mathbf{P}_{\rm intersect}$ on the surface of the tissue. During the training, the ground truth intersection point position was provided by the laser source, while during testing the intersection was estimated solely based on visual information without laser guidance (see Fig.~\ref{fig_problem_define}). 


\subsubsection{Network Architecture.} Unlike the segmentation approach, the intersection point was directly predicted using a regression network. The images fed to the network were `laser off' stereo RGB, but crucially, the intersection point for these images was known \textit{a priori} from the paired `laser on' images. The raw image resolution was $4896\times3680$ but these were binned to $896\times896$. Principal Component Analysis  (PCA)~\cite{mackiewicz1993principal} was used to extract the central axis of the probe and $50$ points were sampled along this axis as an extra input dimension. A network was designed with two branches, one branch for extracting visual features from the image and one branch for learning the features from the sequence of principal points using ResNet~\cite{he2016deep} and Vision Transformer (ViT)~\cite{dosovitskiy2020image} as two backbones. The principal points were learned through a multi-layer perception (MLP) or a long short-term memory (LSTM) network~\cite{hochreiter1997long}. The features from both branches were concatenated and used for regressing the intersection point (see Fig.~\ref{fig_network}). Finally, the whole network is trained end-to-end using the mean square error loss.




\subsection{Implementation}
\subsubsection{Evaluation Metrics.} To evaluate sensing area location errors, Euclidean distance was adopted to measure the error between the predicted intersection points and the ground truth laser points. We reported the mean absolute error, the standard derivation, and the median in pixel units.

\subsubsection{Implementation Details.}
The networks were implemented in PyTorch \cite{paszke2017automatic}, with an input resolution of \(896\times896\) and a batch size of $12$. We partitioned the \textbf{Jerry} dataset into three subsets, the training, validation, and test set, consisting of 800, 200, and 200 images, respectively, and the same for the \textbf{Coffbee} dataset. The learning rate was set to \(10^{-5}\) for the first $300$ epochs, then halved until epoch $400$, and quartered until the end of the training. The model was trained for $700$ epochs using the Adam optimizer on two NVIDIA 2080 Ti GPUs, taking approximately $4$ hours to complete. 

\begin{figure}[t]
\centering
\includegraphics[width=0.99\linewidth]{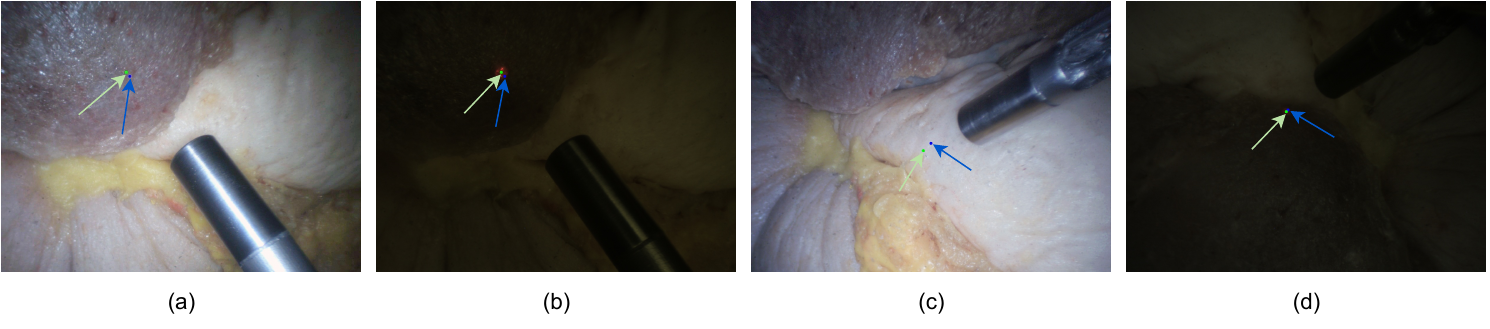}
\vspace{-2ex}
\caption{Qualitative results. (a) and (c) are standard illumination images and (b) and (d) are images with laser on and laparoscopic light off. The predicted intersection point is shown in blue and the green point indicates the ground truth, which are further indicated by arrows for clarity.\\
}
\label{vis}
\end{figure}

\begin{table}[ht]
\begin{minipage}[b]{0.48\linewidth}
\centering
\footnotesize
\setlength{\tabcolsep}{0.3em} 
{\renewcommand{\arraystretch}{1.1}
\begin{tabular}{c|c|c|c|c|c|c}
\hline
\multicolumn{2}{c|}{\textbf{ResNet}}     & \checkmark & \checkmark  & \checkmark  & \checkmark  & \checkmark \\ \hline
\multicolumn{2}{c|}{\textbf{MLP}} &      & \checkmark   & & & \checkmark    \\ \hline
\multicolumn{2}{c|}{\textbf{LSTM}}   &  &   & \checkmark   & &     \\ \hline
\multicolumn{2}{c|}{\textbf{Stereo}} & \checkmark     &      \checkmark     &\checkmark    &  &   \\ \hline
\multicolumn{2}{c|}{\textbf{Mono}} & & &    & \checkmark  & \checkmark        \\ \hline
\hline

\rowcolor[RGB]{230,230,230}\multicolumn{2}{c|}{2D Mean E.} & $73.5$ & $70.5$ &  $73.7$  & $75.6$  & $76.7$        \\ 
\rowcolor[RGB]{230,230,230}\multicolumn{2}{c|}{2D Std.} & $65.1$ & $56.8$ &  $62.1$  & $62.9$  & $64.4$        \\ 
\rowcolor[RGB]{230,230,230}\multicolumn{2}{c|}{2D Median} & $57.5$ & $59.8$ &  $56.9$  & $58.8$  & $68.4$        \\ \hline \hline

\rowcolor[RGB]{193,218,243}\multicolumn{2}{c|}{2D Mean E.} & $63.2$ & $52.9$ &  $62.0$  & $55.8$  & $60.2$        \\ 
\rowcolor[RGB]{193,218,243}\multicolumn{2}{c|}{2D Std.} & $71.4$ & $42.9$ &  $63.4$  & $55.3$  & $42.1$        \\ 
\rowcolor[RGB]{193,218,243}\multicolumn{2}{c|}{2D Median} & $44.9$ & $44.6$ &  $43.4$  & $42.5$  & $52.3$        \\\hline
\rowcolor[RGB]{193,218,243}\multicolumn{2}{c|}{R2 Score} & $0.55$ & $0.82$ &  $0.63$  & $0.73$  & $0.78$        \\ \hline

\rowcolor[RGB]{193,218,243}\multicolumn{2}{c|}{3D Mean E.} & $8.5$ & $7.4$ &  $6.5$  & $6.4$  & $11.2$        \\ 
\rowcolor[RGB]{193,218,243}\multicolumn{2}{c|}{3D Std.} & $15.7$ & $6.7$ &  $6.8$  & $7.1$  & $18.2$        \\ 
\rowcolor[RGB]{193,218,243}\multicolumn{2}{c|}{3D Median} & $4.5$ & $4.6$ &  $4.0$  & $4.3$  & $5.4$        \\\hline

\end{tabular}
}
\vspace{0.25cm}
\caption{Results using ResNet50. Grey color denotes the Jerry dataset and Blue color is for Coffbee dataset (2D errors are in pixels and 3D errors are in mm).}
\label{tab_resnet}
\end{minipage}\hfill
\begin{minipage}[b]{0.5\linewidth}
\centering
\footnotesize
\setlength{\tabcolsep}{0.3em} 
{\renewcommand{\arraystretch}{1.1}
\begin{tabular}{c|c|c|c|c|c|c}
\hline
\multicolumn{2}{c|}{\textbf{ViTNet}}       & \checkmark & \checkmark   & \checkmark  & \checkmark   & \checkmark                                              \\ \hline
\multicolumn{2}{c|}{\textbf{MLP}}     &      & \checkmark     &  &  & \checkmark \\ \hline
\multicolumn{2}{c|}{\textbf{LSTM}}   &   &              & \checkmark   &              &    \\ \hline
\multicolumn{2}{c|}{\textbf{Stereo}} & \checkmark     &  \checkmark &  \checkmark     &  &  \\ \hline
\multicolumn{2}{c|}{\textbf{Mono}} &  &   &  & \checkmark            & \checkmark   \\ \hline
\hline

\rowcolor[RGB]{230,230,230}\multicolumn{2}{c|}{2D Mean E.} & $77.9$ & $92.3$ &  $80.9$  & $87.7$  & $112.1$        \\ 
\rowcolor[RGB]{230,230,230}\multicolumn{2}{c|}{2D Std.} & $69.1$  & $71.0$ &  $67.4$  & $68.6$  & $84.2$        \\ 
\rowcolor[RGB]{230,230,230}\multicolumn{2}{c|}{2D Median} & $59.0$ & $75.0$ &  $64.8$  & $74.9$ & $90.0$        \\ \hline \hline

\rowcolor[RGB]{193,218,243}\multicolumn{2}{c|}{2D Mean E.} & $76.3$ & $75.0$ &  $88.0$  & $56.5$  & $82.7$        \\ 
\rowcolor[RGB]{193,218,243}\multicolumn{2}{c|}{2D Std.} & $69.8$  & $60.6$ &  $83.3$  & $75.8$  & $63.9$        \\ 
\rowcolor[RGB]{193,218,243}\multicolumn{2}{c|}{2D Median} & $59.9$ & $59.6$ &  $68.3$  & $34.5 $ & $69.1$        \\ \hline

\rowcolor[RGB]{193,218,243}\multicolumn{2}{c|}{R2 Score} & $0.58$ & $0.66$ &  $0.33$  & $0.65$ & $0.60$        \\ \hline

\rowcolor[RGB]{193,218,243}\multicolumn{2}{c|}{3D Mean E.} & $7.9$ & $9.1$ &  $11.4$  & $11.6$  & $7.7$        \\ 
\rowcolor[RGB]{193,218,243}\multicolumn{2}{c|}{3D Std.} & $6.9$  & $8.2$ &  $16.7$  & $21.3$  & $7.0$        \\ 
\rowcolor[RGB]{193,218,243}\multicolumn{2}{c|}{3D Median} & $6.0$ & $5.9$ &  $7.1$  & $5.3$ & $6.2 $        \\ \hline

\end{tabular}
}
\vspace{0.25cm}
\caption{Results using ViT. Grey color denotes the Jerry dataset and Blue color is for Coffbee dataset (2D errors are in pixels and 3D errors are in mm).}
\label{tab_vit}
\end{minipage}
\vspace{-0.3cm}
\end{table}

\section{Results}
Quantitative results on the released datasets are shown in Table \ref{tab_resnet} and Table \ref{tab_vit} with different backbones for extracting image features, ResNet and ViT. For the 2D error on two datasets, among the different settings, the combination of ResNet and MLP gave the best performance with a mean error of $70.5$ pixels and a standard deviation of $56.8$. The median error of this setting was $59.8$ pixels while the R2 score was $0.82$ (higher is better for R2 score). Comparing the Table \ref{tab_resnet} and Table \ref{tab_vit}, we found that the ResNet backbone was better than the ViT backbone in the image processing task, while MLP was better than LSTM in probe pose representation. 
ResNet processed the input images as a whole, which was better suited for utilizing the global context of a unified scene composed of the tissue and the probe, compared to the ViT scheme, which treated the whole scene as several patches. 
Similarly, the sampled $50$ principal points on the probe axis were better processed using the simple MLP rather than using a recurrent procedure LSTM. It is worth noting that the results from stereo inputs exceeded those from mono inputs, which can be attributed to the essential 3D information included in the stereo image pairs. 

For the 3D error, the ResNet backbone still gave generally better performance than the ViT backbone while under the ResNet backbone, LSTM and MLP gave competitive results and they are all in sub-milimiter level. We note that the 3D error subjected to the quality of the acquired ground truth depth maps, which had limited resolution and non-uniformly distributed valid data due to hardware constraints. Hence, we used the median depth value of a square area of 5 pixels around the points where depth value was not available. 

Fig.~\ref{vis} shows visualization results of our method using ResNet and MLP. This figure illustrates that our proposed method successfully detected the intersection point using solely standard RGB laparoscopic images as the input. Furthermore, based on the simple design, our method achieved the inference time of $50$ frames per second, making it well-suitable for intraoperative surgery.

\section{Conclusion}
In this work, a new framework for using a laparoscopic drop-in gamma detector in manual or robotic-assisted minimally invasive cancer surgery was presented, where a laser module mock probe was utilized to provide training guidance and the problem of detecting the probe axis-tissue intersection point was transformed to laser point position inference. Both the hardware and software design of the proposed solution were illustrated and two newly acquired datasets were publicly released. Extensive experiments were conducted on various backbones and the best results were achieved using a simple network design, enabling real time inference of the sensing area. We believe that our problem reformulation and dataset release, together with the initial experimental results, will establish a new benchmark for the surgical vision community. 

\bibliographystyle{splncs04}
\bibliography{paper}

\end{document}


%

\title{{Full Supplementary Material with Text}}

\maketitle

This document presents the supplementary materials for the paper ``\textit{Detecting Sensing Area of A Laparoscopic Probe in Minimally Invasive Cancer Surgery}". In Section~\ref{framework}, details of segmentation framework applications are presented with network input and output. In Section~\ref{result}, we describe the results of the segmentation solutions and provide a discussion about them. Future work is outlined in the Section~\ref{future}.

\section{Intersection Detection as Segmentation}
\label{framework}
In this section, we show the details of the application of segmentation networks we adopted for the first attempt, Swin-unet~\cite{cao2023swin} and Transunet~\cite{chen2021transunet}. The input image was croped to $892 \times 892$ same as the input for the regression network and the output was the segmentation map of the laser point, which represented the sensing area of the gamma probe. The central point of the segmented area was calculated from the segmentation map as the intersection point. 

\begin{table}[!h]
\caption{Results using segmentation solutions.}
\centering
\resizebox{0.82\columnwidth}{!}{%
\begin{tabular}{|c|c|ccc|}
\hline
\small
\multirow{2}{*}{Methods} & \multirow{2}{*}{Failure rate} & \multicolumn{3}{c|}{Accuracy on valid frames (pixels)}                                   \\ \cline{3-5} 
                         &                               & \multicolumn{1}{c|}{Mean error} & \multicolumn{1}{c|}{Std.} & Median \\ \hline
Swin-unet~\cite{cao2023swin}   & 34\%                          & \multicolumn{1}{c|}{41.1}          & \multicolumn{1}{c|}{47.0}    & 27.8      \\ \hline
Transunet~\cite{chen2021transunet} & 22\%                          & \multicolumn{1}{c|}{34.5}          & \multicolumn{1}{c|}{40.6}    & 22.2      \\ \hline
\end{tabular}}
\vspace{3ex}

\label{segmentation}
\end{table}

\section{Results and discussion}
\label{result}
The problem of the segmentation solution is the inferior successful rate. Only 66\% and 78\% frames on the test dataset had valid segmented area on the segmentation map and the rest were full of $0$, which noted the background. For the invalid segmentation map, the central point would be NAN, leading to meaningless intersection point. Hence, we only calculated the accuracy on the valid segmentation maps. The best mean error acquired was $34.5$ pixels with standard deviation of $40.6$ and median error of $22.2$ pixels by Transunet~\cite{chen2021transunet}. 

It can be seen from Table \ref{segmentation} that once the frame is successfully segmented, the accuracy of the intersection point calculation would be higher than regression solution. However, the unstable detection rate is a big problem for application in intraoperative surgery. We attributed this to the data over-fitting, which means that $700$ images are not enough to make a segmentation solution work on our task. 

\section{Future work}
\label{future}
In the future, we are going to make use of the stable detection achieved through the regression work, leveraging the high accuracy of the segmentation solution on successfully segmented maps. More data will be collected using the fully sequential and automatic hardware platform introduced in the main paper. Furthermore, our framework will be further validated on the real human \textit{ex-vivo} tissue data.

		 


\bibliographystyle{splncs04}
\bibliography{paper}